\documentclass[apj]{emulateapj}

\usepackage{epsfig}
\usepackage{multirow}
\usepackage{graphicx}
\usepackage{epstopdf}
\usepackage{amsmath, amsthm, amssymb}
\usepackage{natbib}
\usepackage{rotating}
\shorttitle{Chemo-dynamical clustering}
\shortauthors{B. Chen et al. }
\graphicspath{{figs/}}

\begin{document}

\title{Chemo-Dynamical Clustering applied to APOGEE data: 
Re-Discovering Globular Clusters} 

\author{Boquan Chen \altaffilmark{1}, Elena D'Onghia \altaffilmark{1,2}, Stephen A. Pardy
  \altaffilmark{1}, Anna Pasquali \altaffilmark{3}, Clio Bertelli Motta \altaffilmark{3}, Bret Hanlon \altaffilmark{4}, Eva K. Grebel \altaffilmark{3}}

\begin{abstract}
We have developed a novel technique based on a clustering algorithm which searches for kinematically--and--chemically--clustered stars in the APOGEE DR12 Cannon data. As compared to classical chemical tagging, the kinematic information included in our methodology allows us to identify stars that are members of known globular clusters with greater confidence. We apply our algorithm to the entire APOGEE catalog of 150,615 stars whose chemical abundances are derived by the Cannon. Our methodology found anti-correlations between the elements Al and Mg, Na and O, and C and N previously identified in the optical spectra in globular clusters, even though we omit these elements in our algorithm. Our algorithm identifies globular clusters without {\it a priori} knowledge of their locations in the sky. Thus, not only does this technique promise to discover new globular clusters, but it also allows us 
to identify candidate streams of kinematically--and chemically--clustered stars in the Milky Way.

\end{abstract}


\altaffiltext{1}{Department of Astronomy, University of Wisconsin-Madison, 475 N. Charter Street, Madison, WI 53076, USA.
  {e-mail:ebchen.0825@gmail.com}}

\altaffiltext{2}{Center for Computational Astrophysics, Flatiron Institute, 162 Fifth Avenue, New York, NY 10010, USA}
\altaffiltext{3}{Astronomisches Rechen-Institut, Zentrum f\"ur Astronomie der Universit\"at Heidelberg, 
M\"onchhofstrasse 12 - 14, 69120 Heidelberg, Germany}
\altaffiltext{4}{Statistics Department, University of Wisconsin-Madison, 1300 University Avenue, Madison, WI 53076, USA}

\section{Introduction}
\label{sec:intro}

Observations of the stellar content of giant molecular clouds within a few kpcs of the Sun have shown that most stars form in embedded aggregates of rather different sizes and stellar densities 
\citep{Lada2003}. The star formation efficiency and  time scale
of gas dispersal undergone by these stellar groups determine whether or not they can emerge from their
parent cloud as bound clusters. Even so, their internal dynamical evolution and tidal interactions with the Galaxy cause the dissolution of, at least, the less massive clusters whose stars are then released into the 
field population. It is commonly expected that stars formed during the same burst of star formation and in the same cloud ideally would share identical chemical composition, modulo chemical inhomogeneities of the gas within their parent cloud. This is the case for Galactic open clusters, whose single-age stellar populations have been found to be chemically homogeneous \citep{Bovy16}. However, there are still debates about the intrinsic inhomogeneity within a star cluster \citep{Liu2016Hyades}.

The challenge in the decade of large  Milky Way surveys is to find whether the reverse is still valid. The question is to assess whether stars with similar chemical abundances can be identified as members of stellar clusters because of their common birth origin. In
every star, its photospheric element abundance persists over much of its lifetime for the majority of elements. Therefore, stars that are born together in the same cloud or during the same burst of star formation can in principle be identified through their derived photospheric abundances, even after the stars are dispersed and no longer spatially close. The concept of chemical tagging is based on this affirmation. 

Suggested more than a decade ago \citep{Freeman2002, BlandHawthorn2010},
 chemical tagging has motivated many ongoing surveys like APOGEE \citep{Majewski15} and GALAH \citep{DeSilva15}.
Chemical tagging is based on the concept of identifying stars of common birth origin on the basis of
their similar chemical properties without any {\it a priori} information of their positions or velocities. 
While promising to re-build open clusters from stars with similar chemical abundances, chemical tagging has also 
been suggested as a technique to reveal the accretion history of our Galaxy, as stars that form in accreted satellites 
are expected to have distinct chemical properties from the in situ stars that form 
in the central body \citep{Schlaufman2011}.

There are concerns with these assertions and the validity of chemical tagging. First, stellar evolution modifies the initial chemical composition of stars so that stars formed at the same time but now in different evolutionary phases differ in 
their CNO abundances because of convection in their envelope and the occurrence of the first dredge-up \citep{Bertelli17, Salaris2015}. 
In addition, diffusion, i.e.\ the combined effect of 
gravitational settling and radiative acceleration, causes the stellar surface abundances to change during the main-sequence phase. Although for most elements the initial surface abundance is expected to be essentially restored after the main sequence turn-off, through convection during the subgiant and lower red giant phase, the chemical species that are subject to nuclear processing in the stellar interior represent an exception \citep{Bertelli2018, Dotter17}. These processes, whose amplitudes depend on stellar mass and initial chemical composition, limit the precision of chemical tagging. Nevertheless, stars that were born together and are now on the red giant branch should have similar chemistry for most elements, provided that they are born with similar chemical abundances. For these reasons, atomic diffusion might affect the ability to perform chemical tagging with stars in different evolutionary phases, but is not a serious concern when dealing e.g. only with red giants.

Chemical abundances of a star could also be correlated with the time and location of its birth, which poses another limitation on chemical tagging (see \citealt{KrumholzTing2017} for details). Stars that were born about the same time and/or location could share similar abundances even though they are not from the same star formation association.

Secondly, accurate photometry of many Galactic globular clusters has shown the existence of multiple stellar populations \citep{Gratton12} 
with a variation in the abundance of light elements.  
In particular, optical spectra of stars in Galactic globular clusters show anti-correlations between the elements Al and Mg, Na and O \citep{Cohen78,Carretta09a,Carretta09b}, and C and N \citep{Kayser08}. These anti-correlations are interpreted as the chemical signature of a complex star formation history whereby the stellar content of a globular cluster built up through several episodes of star formation. The older stellar generation can also chemically pollute the gas from which the younger stars formed \citep{Bastian15}. 

Furthermore, chemical tagging is a promising technique only if precise measurements of many chemical abundances are taken for each star 
in order to obtain a unique chemical pattern. Hence, to be effective, chemical tagging needs to operate in a multi-dimensional chemical space. This requirement motivated new surveys to observe abundances for several elements, e.g. APOGEE \citep{Smith2013} and GALAH \citep{DeSilva15}. 

Concerning the precisions of chemical abundances, the signal-to-noise ratios at relevant locations in spectrum space determine whether high-precision tags will be delivered \citep{Hogg16}. In particular, the Cannon \citep{Ness15}, a data-driven model, is an effective technique for inferring chemical abundances from lower signal-to-noise stars.

Other caveats about the applicability of chemical tagging concern the stellar disk of our Galaxy. Open stellar clusters are expected to be homogeneous in chemical abundance when they are formed. However, in the disk, most of the stars are born unbound or in unbound associations, or in open clusters that disperse rapidly. It is also important to note that unbound clusters can be chemically homogeneous \citep{Feng2014, KrumholzTing2017}. The average life time of an open cluster in the Galactic disk is $\sim$330 Myr, and estimates suggest that up to 40\% of the disk stars were once born in open clusters \citep{Piskunov08, Lewis2015}. \citet{Pancino2018} found peculiar abundance patterns and Na-O anti-correlations in stars of intermediate-age open clusters in the Milky Way.

Moving groups -- associations of stars with coherent velocities -- are either dissolved star clusters, dissolved dwarf galaxies \citep{Meza}, or resonances induced by the presence of spiral arms or the bar \citep{Dehnen,Quillen,BovyHogg}. Most of the moving groups in the solar neighborhood are shown to be chemically inhomogeneous and contain stars of different ages, which leads to the interpretation of having been caused by dynamical perturbations \citep{Antoja2008, Antoja2012, 2011Hyades, Ramya2012}.

Other processes in the stellar disk operate to exchange the angular momentum and energy of individual stars with neighboring stars, such as radial migration \citep{Sellwood02,Minchev11,VeraCiro1,VeraCiro2,DanielWyse}. Indeed, old stellar populations show a wide range of abundances at any given Galactocentric distance, which might be caused by radial migration \citep{Haywood08}.

Despite these limitations, our algorithm analyzes the entire APOGEE sample. Our findings suggest that chemical tagging alone has a limited ability to retrieve stars that are members of known globular clusters. Instead, we show that the chemical properties when combined with the stellar kinematic information, e.g. radial velocity, are promising in re-building the known globular clusters in the sky.

The data sample is presented in Section~\ref{sec:data}.
The various algorithms used for chemical tagging are 
described in Sections ~\ref{sec:Kmeans} and ~\ref{sec:DBSCAN}. We detail a
new algorithm that combines chemical and kinematic information in  
Section~\ref{sec:SNN}. Section~\ref{sec:results} shows the results when this algorithm is applied to the APOGEE DR12 Cannon data, and  Section~4 summarizes 
the main results of this study.

\section{Properties of the Cannon Sample}
\label{sec:data}

In this study, we aim to re-build globular clusters in the Milky Way. Instead of using the publicly available APOGEE DR12 data \citep{Holtzman15}, 
we use stellar parameters derived by a data-driven method called the Cannon \citep{Casey16}. The Cannon is able to derive chemical abundances for many 
stars with low signal-to-noise ratios in APOGEE DR12 and thus allows us to analyze the chemical abundances of more stars, including many known members. Our sample contains 150,516 stars that have valid values for fifteen chemical elements: C, N, O, Na, Mg, Al, Si, S, K, Ca, Ti, V, Mn, Fe, and Ni.

As discussed in previous studies, chemical tagging is better performed in the chemical space normalized by iron ([X/Fe], with X being a chemical element), instead of the space normalized by hydrogen ([X/H]) \citep{Ting12}. [X/H] values strongly correlate with each
other, making it harder to observe the subtle variations among individual clusters in the chemical space. Therefore, in this study, we consider an abundance space spanned by iron [Fe/H] and the other elements [X/Fe]. 

Next, we clarify the dimensionality of the chemical space we consider. 
While other studies have discarded Ti and V because the abundances of these elements might not be reliable yet in DR12, we opt to keep these two elements in our algorithm. The biases in these elements would not affect the performance of our algorithm, as long as they have high precision. We discard C and N because they are expected to evolve through stellar evolution due to the post main sequence dredge--up 
(e.g. \citealt{Masseron15, Ventura16}). Thus, it is complicated to relate C and N to their initial abundances when the stars formed. Discarding C, N and the other four elements expected to anti-correlate in globular clusters, i.e.\ Na, O, Mg, Al, we consider nine elements in this study, namely Si, S, K, Ca, Ti, V, Mn, Fe, Ni. We note that our results do not change significantly even if all elements are included in the algorithms. 

APOGEE observed 20 open and globular clusters for calibration purposes. The sample of clusters that were targeted for calibration spanned a wide range of 
metallicities according to the previous literature. Observed targets were considered members according to their previously published chemical abundances, their radial velocities, or their proper motion measurements (for more details, see the discussion reported in \citealt{Meszaros13,Zasowski13}. We denote the rest of the stars as background stars, which do not have membership information.) 

The following globular clusters are targeted by the APOGEE and included in our sample: M107, M13, M15, M2, M3, M5, M53, M92, NGC 5466, as listed in Table
1. Column two quotes the number of known members for each cluster. The following columns list the recovery rate and false positive rates of globular clusters recovered by the various clustering algorithms explored and applied in the next sections of this study. The criteria for categorizing a known member as ``recovered'' are discussed in the beginning of Section 3.

We note that only less than 15 \% of stars in APOGEE have [Fe/H] $<$ -1. One could believe that the halo stars observed in APOGEE are almost exclusively members of globular clusters in the halo. For this reason, we display the background stars divided into two separate groups, [Fe/H] $<$ -1 and [Fe/H] $\geq$ -1, in our clustering results in order to verify that our results are not caused by the way in which the APOGEE samples the halo.

\subsection{Al-Mg and Na-O anti-correlations in the DR12}

Globular clusters are characterized by stars with significant anti-correlations among
[Al/Fe]-[Mg/Fe] and [Na/Fe]-[O/Fe] but overall homogeneous Fe \citep{Carretta09a,Carretta09b}. Before
describing the algorithms that perform the clustering analysis on the data set, we will investigate the presence of these anti-correlations in the DR12 Cannon data for the globular clusters M107, M13, M15, M2, M3, M5, M53, M92, and NGC 5466.

\begin{table*}[htbp]
   \centering
	\caption{GLOBULAR CLUSTERS in APOGEE \\
	Recovery and False Positive Rates for Globular Clusters} 
   \begin{tabular}{@{} llcccc @{}} 
      \hline
      Name    & Alternative & Members & K-Means$^{2}$ &  DBSCAN$^{2}$ & SNN$^{3}$  \\
              & designation & Number  &   R/FP[\%]$^{1}$  & R/FP[\%]$^{1}$ & R/FP[\%]$^{1}$\\
      \hline

      NGC 6171 & M107  &  18  & ...   & ...  & ...   \\
      NGC 6205 & M13   &  71  & 26.7/66.7 & 23.9/48.5$^{4}$ & 95.8/50.7    \\
      NGC 7078 & M15   &  11  & ...   & ...  & ...      \\
      NGC 7089 & M2    &  19  & ...   & ...  & ...    \\
      NGC 5272 & M3    &  73  & ...   & ...  & 34.2/35.9    \\
      NGC 5904 & M5    &  103 & ...   & 18.4/9.5 & ...   \\
      NGC 5024 & M53   &  16  & ...   & ...  & ...   \\
      NGC 6341 & M92   &  48  & ...   & ...  & 64.6/45.6$^{4}$  \\
      NGC 5466 & ...   &   8  & ...   & ...  & 75.0/14.3  \\
      \hline
       \end{tabular}
\\
$^{1}$ R stands for recovery rate and FP stands for false positive rate. \\
$^{2}$ Clustering in the chemical space. \\
$^{3}$ Clustering in the chemistry-velocity space. \\
$^{4}$ Members recovered in multiple groups. \\
$\ \ \ \ \ \ \ $
\label{tab:table1}
\end{table*}

\begin{figure*}[htb]
   \includegraphics [width=1.0 \textwidth]{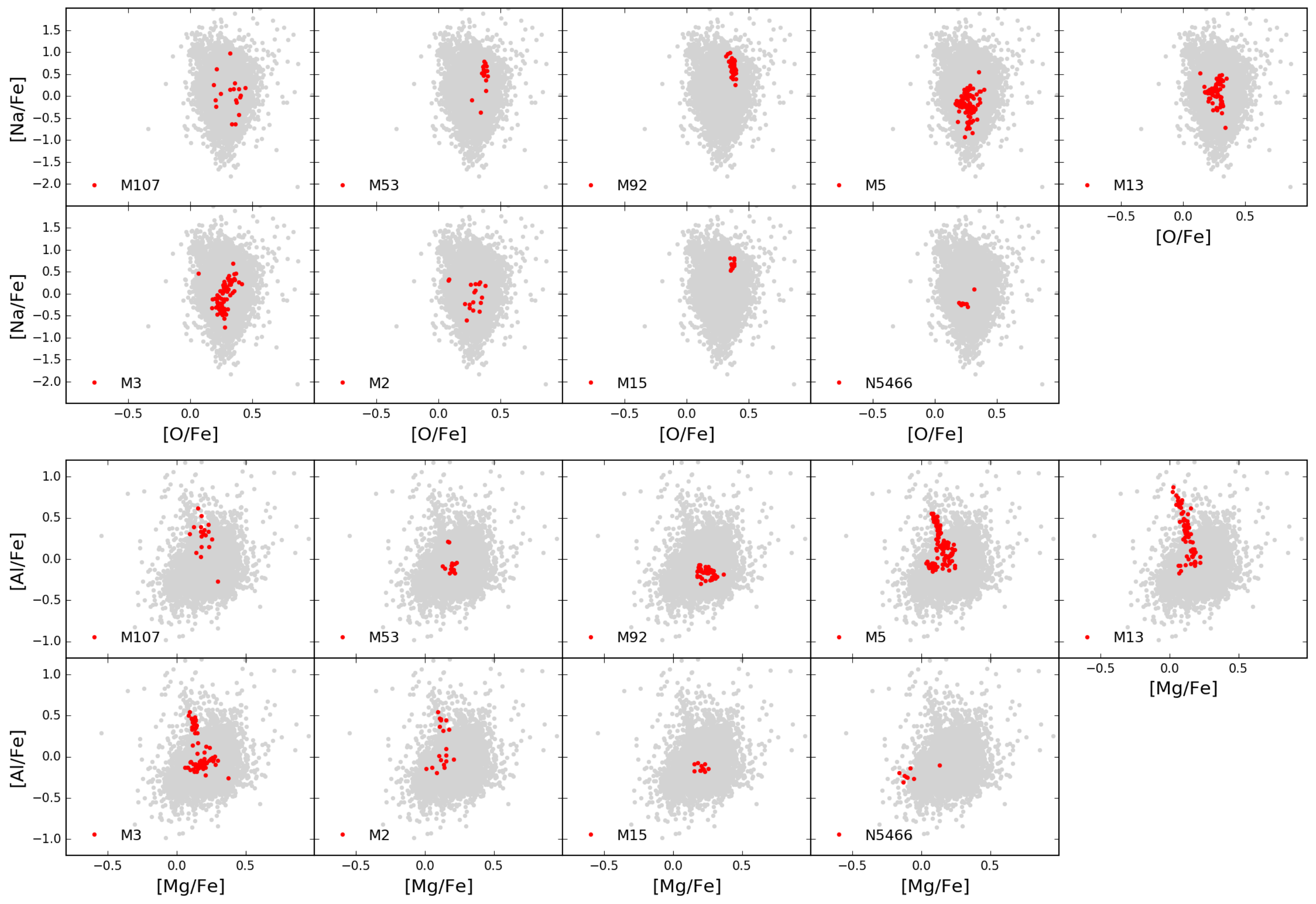}
  \caption{[Na/Fe]-[O/Fe] and [Mg/Fe]-[Al/Fe] anti-correlation for a total of 367  
individual giant stars (red dots) in the globular clusters identified in DR12 using the Cannon over all stars in the sample with [Fe/H] $<$ -1 (gray dots).
{\it Top Panels.} [Na/Fe] measurements of stars are plotted against [O/Fe] abundances of the same stars. There is a lack of anti-correlation between [Na/Fe]-[O/Fe] in the APOGEE DR12 data as compared to the anti-correlation found in the optical data that characterize globular clusters. {\it Bottom Panels.} The [Al/Fe]-[Mg/Fe] anti-correlation is shown for the same stars in the DR12 using the Cannon. Note that some background stars exhibit an anti-correlation near [Al/Fe] = 0.5 dex. These background stars overlap with M13 members. }
\label{fig:AlMg}
\end{figure*}

Fig.~\ref{fig:AlMg} plots [Na/Fe]-[O/Fe] (top panels) and [Al/Fe]-[Mg/Fe](bottom panels) for stars in the DR12 Cannon that are members of the observed globular clusters. M107, M5, M13, M3, and M2 show evidence of anti-correlation in the [Al/Fe]-[Mg/Fe] plots. However, [Na/Fe] does not anti-correlate with [O/Fe] in the DR12 Cannon, except for a weak anti-correlation displayed in M92. One possible reason for the lack of anti-correlation between Na and O is that Na has relatively few and weak features, as discussed in \citet{Holtzman15}. \citet{Fernandez2017} reported peculiar abundance patterns in 11 red giant stars that show Al and N enhancements with C and Mg depletions. These patterns are similar to those exhibited in second-generation stars of globular clusters. Similarly, \citet{MartellGrebel2010} and \citet{Martell2011} found globular-cluster-like C-N abundance variations in a subset of halo field stars, suggesting that these stars had a globular cluster origin.

In Fig.~\ref{fig:AlMg}, we also found an anti-correlation between [Al/Fe] and [Mg/Fe] in some background stars with [Fe/H] $<$ -1. These background stars correspond to members of the M13, and possibly M2, M3, and M5 globular clusters in the chemical space. Ideally, a successful chemical tagging algorithm should be able to recover this over-dense structure. However, as we noted before, our results remained largely unchanged when all elements were included in the algorithms. Even though these background stars appear as a over-dense region when projected onto [Al/Fe] and [Mg/Fe], conditioned on their [Fe/H] values, our ability to recover them with chemical tagging is limited. These stars do not share similar radial velocities and can potentially be separated with chemo-dynamical tagging.

M5, M15 and M107 are the only globular clusters that are observed by the APOGEE and also present in the sample studied by
\citet{Carretta09a} at optical wavelengths. When comparing our [Al/Fe]-[Mg/Fe] plots with those of \citet{Carretta09a}, 
we noticed that the DR12 Cannon [Mg/Fe] abundances are systematically lower by $\sim$0.2 dex on average than those in 
\citet{Carretta09a}, while the ranges in [Al/Fe] are consistent with each other. Moreover, while the data collected by \citet{Carretta09a} for red giant stars in these clusters do not show any kind of correlation,
we see a clear [Al/Fe]-[Mg/Fe] anti-correlation in M5 and possibly in M107, but nothing in M15 in our plots.

We also compared the DR12 Cannon distributions with those obtained by  \citet{Meszaros15}. Both data sets cover the same abundance ranges for each cluster except for NGC 5466 whose DR12 Cannon [Mg/Fe] abundances are shifted towards lower values by $\sim$0.15 dex. Both \citet{Carretta09a} and \citet{Meszaros15} confirm an [Al/Fe]-[Mg/Fe] anti-correlation in globular clusters. The Cannon data indicate a possible [Al/Fe]-[Mg/Fe] anti-correlation in M107 which is not seen in \citet{Meszaros15}, but do not exhibit any [Al/Fe]-[Mg/Fe] anti-correlation in NGC 5466, M15, M53 and M92, otherwise detected in \citet{Meszaros15}.
\par

\section{Searching for Globular Clusters}

In this section, we describe two popular chemical-space clustering algorithms, namely K-Means and Density Based Spatial Clustering of Application with Noise (DBSCAN) \citep{Ester96}, discuss their advantages and disadvantages, 
and propose a new chemo-dynamical
algorithm based on the Shared Nearest Neighbors Clustering (SNN) \citep{ertoz03}. Our
data set includes chemical abundances from the APOGEE DR12 Cannon and globular cluster membership information from \citet{Meszaros13}. The membership information was discarded when we performed the algorithms, and was only considered afterward to test the validity of our results.
Each of our three algorithms produced groupings of stars in $N$-dimensional (in the case of the SNN, $N+1$) space, with $N$ being the number of chemical elements used. 

A successful algorithm produces star groupings that have two properties: a low false positive rate, and a high recovery rate. The false positive rate is determined by how many background stars (stars that are not members of any globular cluster) and members of other clusters appear in a group that is supposedly composed of confirmed members of a single globular cluster. If all stars in a group are verified members of the same globular cluster, the group would have a false positive rate of 0. A group of stars from the clustering results is considered a ``valid'' group only if it contains members from the same globular cluster and has a false positive rate of less than 85\%. This is an arbitrary threshold to help us compare results across different algorithms. The recovery rate is determined by what fraction of confirmed members from a certain globular cluster appears in all valid star groupings. Similarly, we combine all valid groups to calculate the false positive rate for a specific globular cluster. Fig. \ref{fig:Kmeansf}, \ref{fig:DBSCANf} and \ref{fig:SNNf2} only circle members ``recovered'' in valid groupings. The recovery rates and false positive rates for the globular clusters are summarized in Table \ref{tab:table1}.

\subsection{Chemical-space clustering with K-Means}
\label{sec:Kmeans}

As a first step we used the K-Means algorithm in order to re-build globular
clusters which are expected to appear as over-dense or clustered regions 
in the $N$-dimensional chemical space. The K-Means algorithm is a clustering algorithm in data mining that has been applied for chemical tagging in previous work 
\citep{Hogg16}. The algorithm aims to partition data points into K groups, each of which contains a core position (not necessarily a data point) and the data points nearest to it.

\begin{figure*}[htb]
  \begin{center}
   \includegraphics [width=1.0 \textwidth]{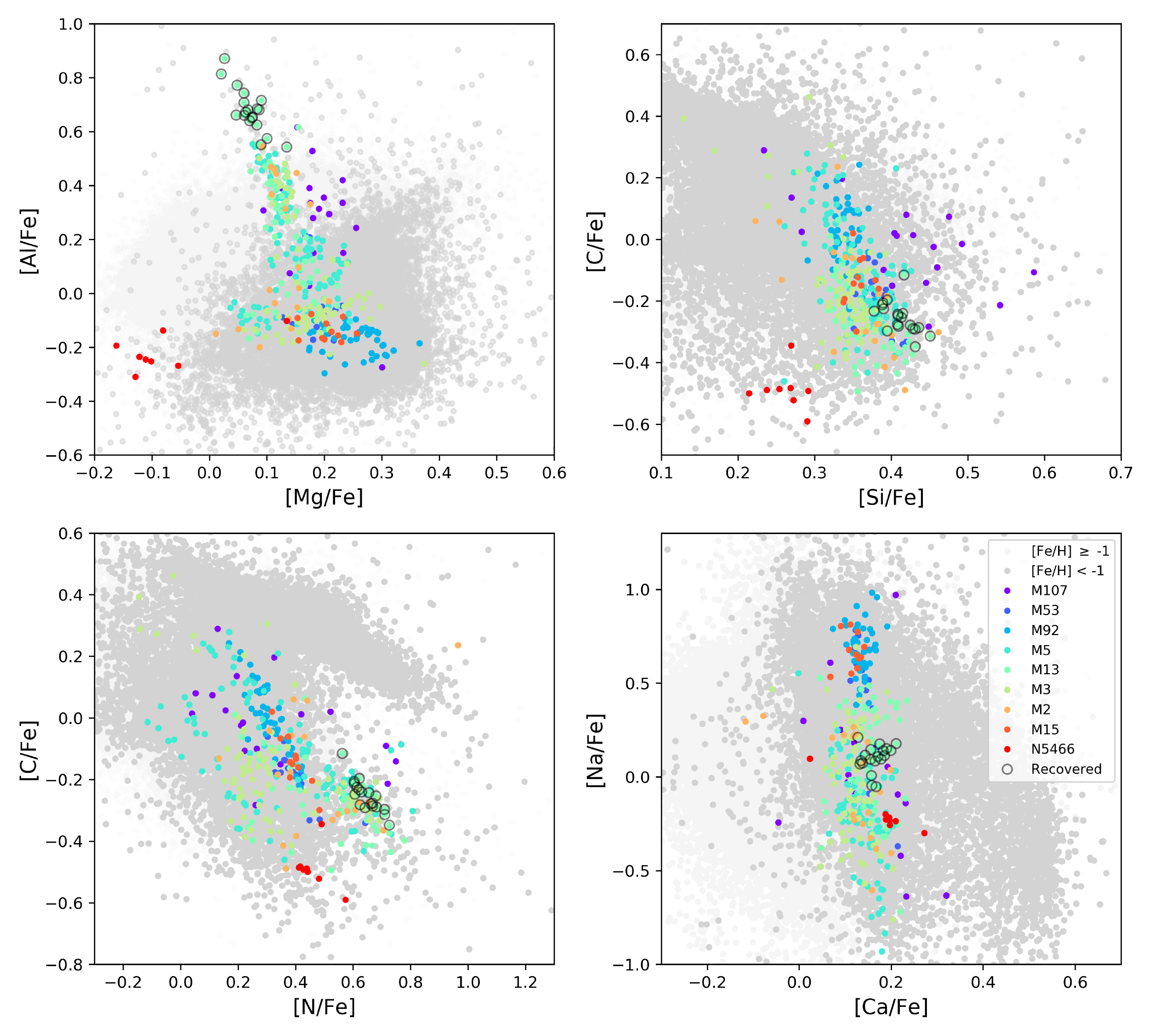}
  \caption{Four two-dimensional projections of the abundance distribution and clustering results using the K-Means algorithm. The algorithm described in Section \ref{sec:Kmeans} performs clustering analysis in the N-dimensional chemical space. Plots are referred to Fe for familiarity reasons. Background stars in the APOGEE (gray for [Fe/H] $<$ -1 and light gray for [Fe/H] $\geq$ -1) and known members of globular clusters (large colored dots) are plotted. The members recovered  by chemical tagging with K-Means are also plotted (open black circles). The majority of known members are not recovered (see Table \ref{tab:table1} for details).}
\label{fig:Kmeansf}
\end{center}
\end{figure*}

\par
Here we explain how K-Means operates in the $N$-dimensional abundance spaces following \citet{Hogg16}:
\begin{itemize} 
\item [1.] \ Choose a value of K that defines the number of groups in the N-dimensional space. 
\item [2.] \ An initial position is chosen to be the center for each group; 
\item [3.] \ Assign each star to the group with the closest center;  
\item [4.] \ Re-center each group to the centroid of its stars;  
\item [5.] \ Iterate the procedure in steps 3 and 4 until convergence has been reached.
\end{itemize}

\noindent

As discussed before, we chose to work in the iron-normalized abundance space. Compared to \citet{Hogg16}, we discarded 6 abundances according to anti-correlations found in [Al/Fe]-[Mg/Fe] and [Na/Fe]-[O/Fe], discussed in \cite{Carretta09a}, as well as [C/Fe] and [N/Fe]. K-Means is usually run several times, with different K to determine the optimal number K. However, we only show the results with K=512 here. Other choices of K do not significantly improve the results.

While widely popular, K-Means faces some limitations. In its initial and basic implementation, it uses Euclidean distances in the N--dimensional space, and assumes a spherical shape and a fixed size of potential clusters in the data. 
In particular, the K-Means algorithm assigns every star to one group. This limits its applicability, since background noise stars are forced by the algorithm to always be assigned to a group. This limitation also means that the over-densities recovered by K-Means in the chemical space will have many false positives and are not guaranteed to be true clusters. We find that for low values of K, the algorithm tends to combine separate known clusters into a mono-abundance group. At high K values, K-Means tends to divide a known cluster into separate groups. 

\begin{figure*}[htb]
	\begin{center}
		\includegraphics[width=1.0\textwidth]{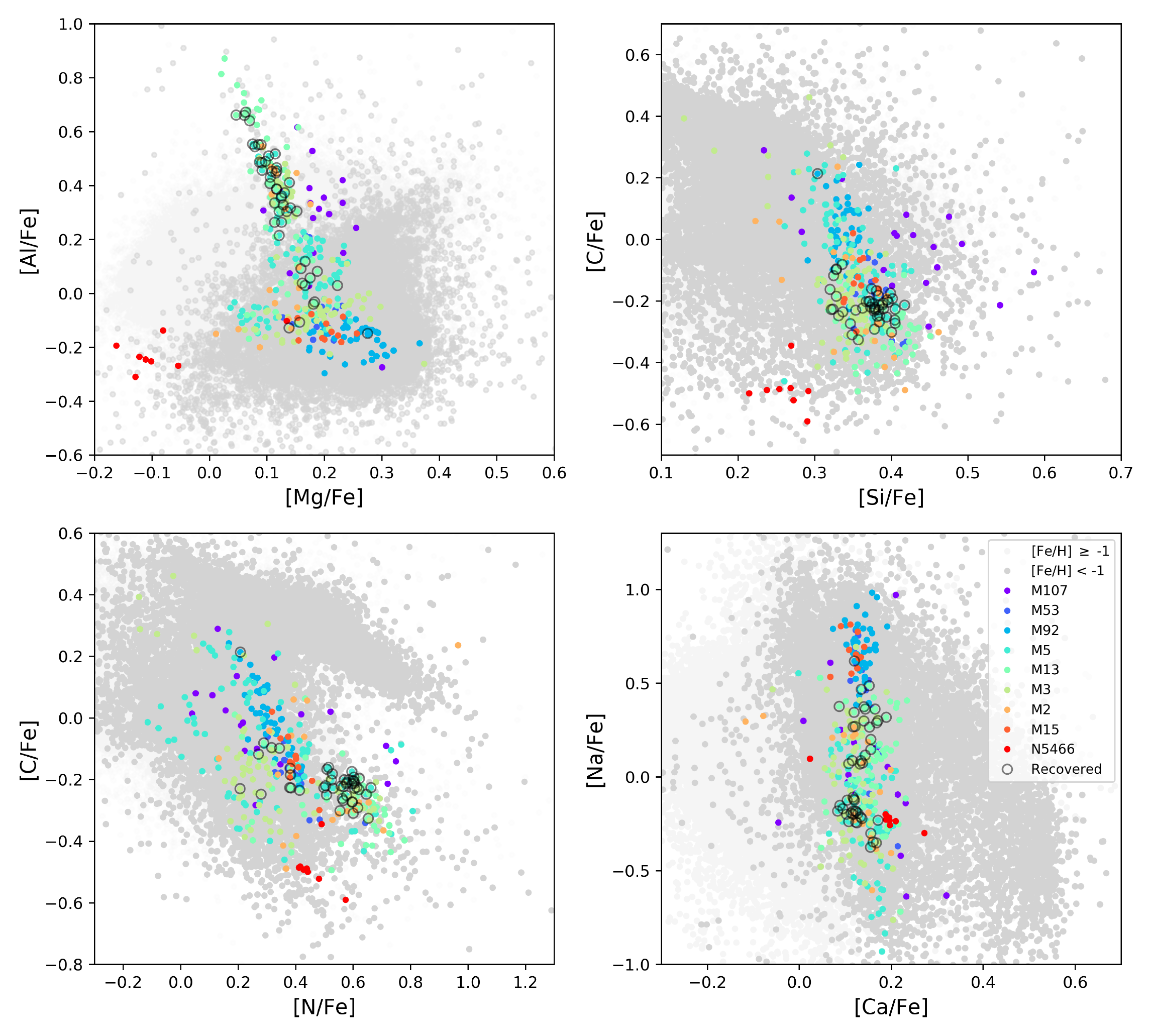}
		\caption{Four two-dimensional projections of the abundance distribution and clustering results using DBSCAN. 
			The clustering algorithm described in Section \ref{sec:DBSCAN}
			performs clustering analysis in the N-dimensional chemical space. Similar to Fig.~\ref{fig:Kmeansf}, background stars in the APOGEE (gray dots) and known members of globular clusters (large colored dots) are plotted. Members recovered by chemical tagging with DBSCAN are also plotted (open black circles). Compared to K-Means, DBSCAN has slightly lower recovery rate but better false positive rates for M13. DBSCAN also identified some M5 members. (see Table \ref{tab:table1} for details). }
		\label{fig:DBSCANf}
	\end{center}
\end{figure*}

Fig.~\ref{fig:Kmeansf} displays the two-dimensional projections of our sample in the chemical space and the results from the K-Means algorithm with K=512. Colored symbols represent the APOGEE stars that are known members of globular clusters (with different colors representing the different clusters targeted in the data set). The gray symbols are the background stars (not members of any globular cluster). The black encircled symbols are the members recovered by K-Means.  This figure demonstrates that stars appearing similar in chemical properties according to the K-Means algorithm are rarely members of any known globular clusters. Members of the M13 globular cluster are recovered by the algorithm with a recovery rate of 26.7\% and a false positive rate of 66.7\% in a single K-Means group. None of the members in the other globular clusters is identified in our sample with K-Means.

\subsection{Chemical Space Clusters with Density Based Spatial Clustering of Application with Noise}
\label{sec:DBSCAN}

For a general data set, if we know the number of disjoint groups and that data points  within each group are normally distributed in our data, the underlying structures of the data set can usually be determined by using K-Means. However, in many cases the groups that characterize the data set do not have a spherical shape, or their number is unknown. In these cases, a more flexible algorithm is required to re-build globular clusters, e.g. DBSCAN, to account for more complex shapes and sizes of clusters.The basic premise behind the algorithm is that each group is expected to be defined by a probability distribution and the probability of picking a star near the central core of the distribution should be much higher than at the edge. If two groups overlap slightly, the overlap will be within the outer edges of the distributions where the probability is negligibly low. Hence, DBSCAN ignores the data points that are sparse and focuses on connecting the dense regions, which should be the central cores of potential groups.

Two parameters have to be chosen for this algorithm:
the maximum radius of the neighborhood from a core
position, $\epsilon$, which determines how large an area around
each data point is considered; the minimum number of data points $K$ inside the $\epsilon$ neighborhood for a point to be considered a core point, MinPts. A data point is a core point if there are at least $K$ other data points that are at
most distance $\epsilon$ away from it. The DBSCAN algorithm
works by connecting all pairs of data points such that 
the two data points (stars) are at most $\epsilon$ away and at least one of them is a core point.

This algorithm has some limitations as well. The results rely on the
choice of parameters $\epsilon$ and $K$. The results can be improved by 
running the algorithm multiple times on data with different parameters. The more important caveat is that it does not perform effectively on data sets where  
the overlap of the two groups extends more than just along their periphery, or where distinct groups have different internal densities. In the latter case, DBSCAN might interpret the less dense group as a collection of outliers. However, the definition of the periphery changes with different norms. Therefore, it is possible to overcome this limitation by adopting a different norm. Consistent with \cite{Mitschang13}, we use the Manhattan distance, or $L_1$ norm, as the metric for implementing chemical tagging with DBSCAN.

Fig.~\ref{fig:DBSCANf} illustrates the two-dimensional projections of the chemical space for our sample and the recovered members according to the DBSCAN algorithm with $\epsilon = 0.4$ and $K=6$. We found that the recovery rate quickly diminishes when we deviate from these values. As compared to chemical tagging performed with K-Means, DBSCAN recovers M13 at a similar rate (23.9 \%)  and 18.4\% of M5 members. However, the M13 members are recovered in four DBSCAN groups, compared to just one group for K-Means. 

\subsection{Chemo-Dynamical Clustering}
\label{sec:SNN}
\noindent

\begin{figure*}[htb]
  \begin{center}
   \includegraphics[width=1.0\textwidth]{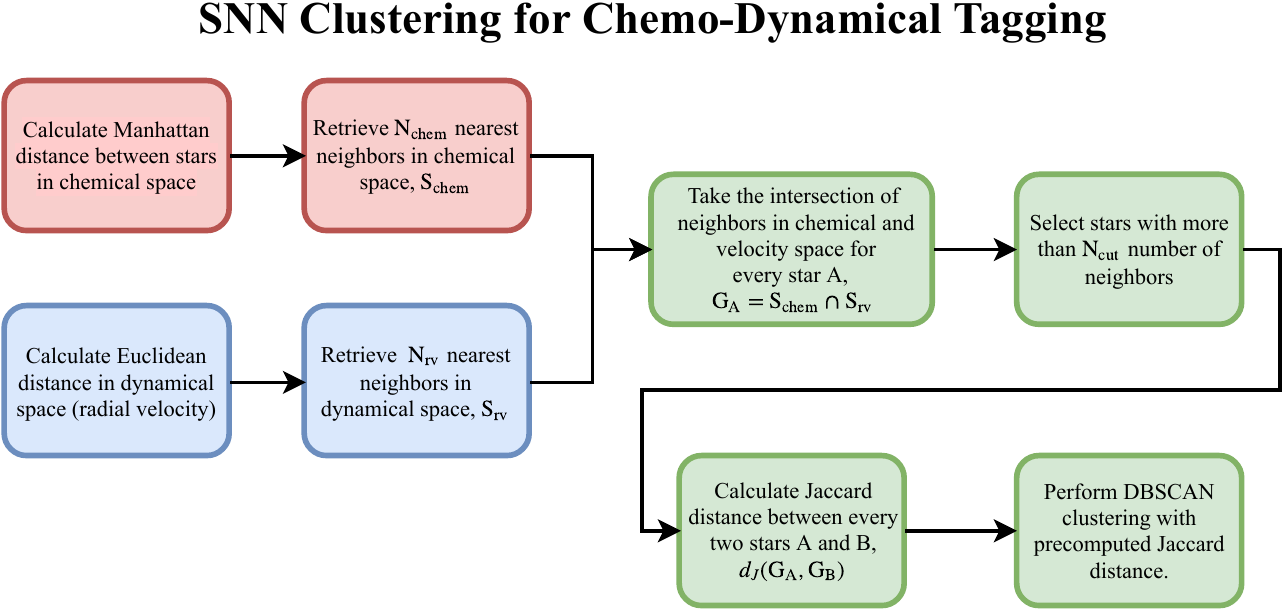}
  \caption{A schematic view of the SNN algorithm with the sequence of the steps used. }
\label{fig:SNNf1}
\end{center}
\end{figure*}

Previous sections showed that the process of identifying groups in chemical
space can retrieve at most 30\% of true members in known globular clusters. {\it This result holds independently of the algorithm adopted to identify clusters in the chemical space.} This is likely due to insufficient precisions of chemical abundances. We found that radial velocities of members from distinct globular clusters are almost indistinguishable in several cases. Thus, radial velocity alone is likely not sufficient to recover individual globular clusters either.

\begin{figure*}[tb]
  \begin{center}
   \includegraphics[width=1.0\textwidth]{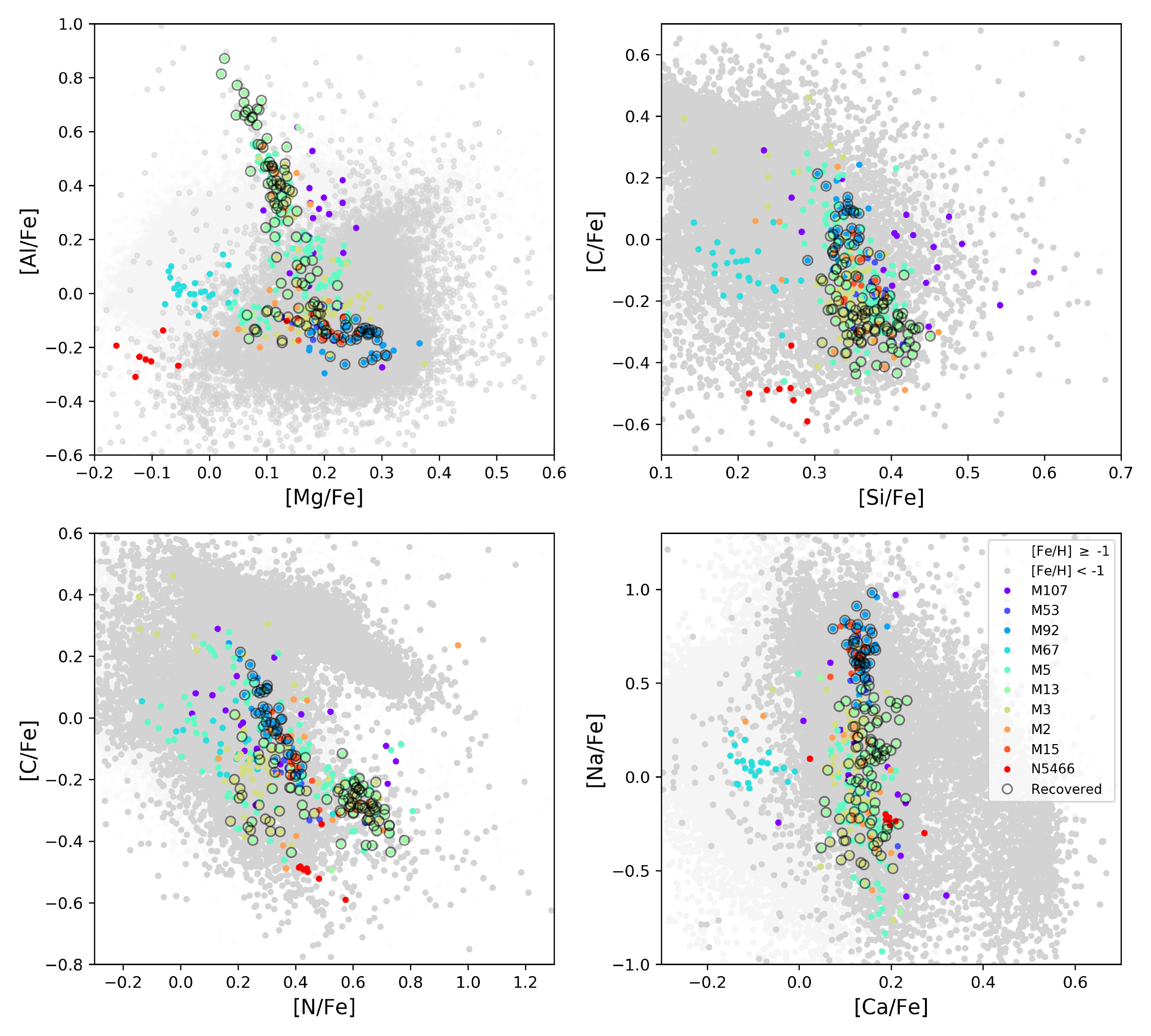}
  \caption{Four two-dimensional projections of the abundance distribution and clustering results using the new SNN algorithm described in Section \ref{sec:SNN}. The clustering analysis is performed in the N+1-dimensional chemical and radial velocity space. Similar to Fig.~\ref{fig:Kmeansf} and \ref{fig:DBSCANf}, background stars in the APOGEE (gray dots) and known members of globular clusters (large colored dots) are plotted. Members recovered by chemistry--velocity tagging with SNN are also plotted (open black circles). The recovery rate of known members is significantly improved over our implementation of DBSCAN and K-Means (see Table \ref{tab:table1} for details).}
\label{fig:SNNf2}
\end{center}
\end{figure*}

As the next step, we add the kinematic information, i.e.\ radial velocity, to the chemical space and propose a new algorithm, SNN. We used AstroPy \citep{astropy} and various SciPy packages \citep{scipy}, including Numpy \citep{numpy}, and 
Scikit-learn \citep{sklearn}, to build our algorithm. We applied this algorithm to our sample and identified star clusters in the chemistry-velocity space.

All clustering techniques rely on assigning stars based on some distance metric. One problem with picking a distance metric is normalization 
when two spaces with different ranges are considered (e.g.\ the chemical space and radial velocity space). {\it One of the main advantages of our algorithm is that it identifies clusters in the chemistry--velocity space by taking the intersection of the nearest neighbors in the chemical space and radial velocity space.} This feature avoids normalization between chemical abundances and radial velocity and guarantees that the final nearest neighbors are neighbors in both the chemical and velocity space. 

The algorithm operates as follows. As suggested in \cite{Mitschang13}, for each set of elements $\{X_i|i = 1,2,..., N\}$,  the distance in the chemical
space between two stars, $A$ and $B$, is computed as the Manhattan distance defined below: 
\begin{equation}
d_{chem} = \sum_{i=1}^N |X_{iA} - X_{iB}| 
\end{equation}

\noindent where $X_i$ = [Fe/H],[K/Fe],[S/Fe],[Si/Fe],[Ca/Fe],[Mn/Fe]$...$. 

The distance between the radial velocities of two stars is a simple Euclidean distance:
\begin{equation}
 d_{rv} = \sqrt{(v_{A} - v_{B})^2}
\end{equation}

For every star, a list of nearest  neighbors in the chemical space and nearest neighbors in the radial velocity space are respectively found. The neighbors for a particular star in the chemo-dynamical space are the intersection of neighbors in the chemical space and velocity space. 
 
Then, the Jaccard distance between each pair of stars is computed by:
\begin{equation}
d_J(G_A, G_B) = 1- \frac{|G_A \cap G_B|}{|G_A \cup G_B|}, 
\end{equation}

\noindent where $G_A$ and $G_B$ are the lists of neighbors for star A and B, respectively, $\cap$ is the intersection, $\cup$ is the union of two sets of neighbors, and the absolute values represent the cardinality of a set. The chemistry-velocity clustering analysis is performed next using the Jaccard distance values and the DBSCAN algorithm.

Our approach has two major advantages. First, as mentioned earlier, we
are able to utilize radial velocity and chemical abundances together without compromising either. Each neighbor of a star is a neighbor in both chemical and velocity space. Second, unlike K-Means and DBSCAN, the SNN algorithm does not make assumptions about the sizes, shapes and densities of clusters \citep{ertoz03}.

A schematic view of our methodology is illustrated in Fig.~\ref{fig:SNNf1} and summarized as follows:

\begin{itemize} 
\item [1.] \ Calculate for each pair of stars the distance $d_{chem}$ in the N-dimensional chemical space (N=9);  
\item [2.] \ Calculate for each pair of stars the distance $d_{rv}$ in radial velocity; 
\item [3.] \ Find the $N_{chem}$ nearest neighbor stars in the chemical space; 
\item [4.] \ Find the $N_{rv}$ nearest neighbor stars in the radial velocity space;
\item [5.] \ Take the intersection of neighbors in the chemical space and velocity space;
\item [6.] \ Remove stars with less than $N_{cut}$ number of neighbors;
\item [7.] \ Calculate the Jaccard distance between every pair of stars;
\item [8.] \ Perform DBSCAN clustering with the precomputed Jaccard distances;
\end{itemize}

Based on exploratory analysis, we chose $N_{chem}$=700 and $N_{rv}$=500. We picked
both numbers to ensure that most known members of globular clusters have some neighbors in the chemistry-velocity space. Given the choices of $N_{chem}$ and $N_{rv}$, we found that 96\% of globular cluster members have at least one neighbor in the chemistry-velocity space. We will not have this information in a true blind search scenario. We instead have to rely on other methods to choose parameters. The numbers $N_{chem}$ and $N_{rv}$ should also not be set too high in order not to cause members from separate globular clusters to share neighbors. We choose a higher number for $N_{chem}$ because globular clusters in the DR12 Cannon are more homogeneous in radial velocity than in chemical abundances. The parameters used for DBSCAN are $\epsilon = 0.35$ and MinPts = 7. These parameters indicate that a core point should have at least seven neighbors that share at least 65\% common neighbors out of all neighbors. If a star has less than $N_{cut}$ number of neighbors, it will be removed from our sample before we perform DBSCAN. We used $N_{cut}$ to eliminate stars without any neighbors in our implementation, or equivalently $N_{cut}=1$.

\subsection{Re-discovering M13}
\label{sec:results}

We show next that our methodology based on the SNN algorithm is promising in discovering new globular clusters or detecting debris stars related to the globular clusters orbiting in our Galaxy.

We applied the SNN algorithm to the DR12 Cannon sample, searching for clusters in the chemistry-velocity space. The groups identified with the SNN are then matched in equatorial coordinates with the catalogue of Galactic globular clusters of \citet{Harris96}. The following globular clusters have been identified: M13, M3, M92, and NGC5466, for which we retrieved the corresponding tidal radii as listed in \citet{Harris96}. However, the membership information only comes from APOGEE, consistent to our previous discussion. Similar to Fig.~\ref{fig:Kmeansf} and \ref{fig:DBSCANf}, we plot the recovered stars from SNN in Fig.~\ref{fig:SNNf2}. Although we highlight members recovered in groups with a false positive rate lower than 85\%, the exact false positive rates are shown in Table \ref{tab:table1}.

Our implementation of the SNN algorithm took $\sim$ 20 minutes. Although it is slower than DBSCAN alone ($\sim$ 5 mins), it is significantly faster than the standard implementation of K-Means in scikit-learn ($\sim$ 75 minutes).

Once the groupings were returned with their stars identified as candidate members, we displayed for each group the two-dimensional projections in the chemical space. An example is illustrated by group 32 identified by the SNN algorithm as a cluster of stars in the chemistry-velocity space. Although it does not overlap perfectly with a globular cluster, the SNN group 32 is dominated by members of M13, which are indicated by open circles in Fig.~\ref{fig:SNNf2}.

\begin{figure}[tb]
  \begin{center}
   \includegraphics[width=0.5 \textwidth]{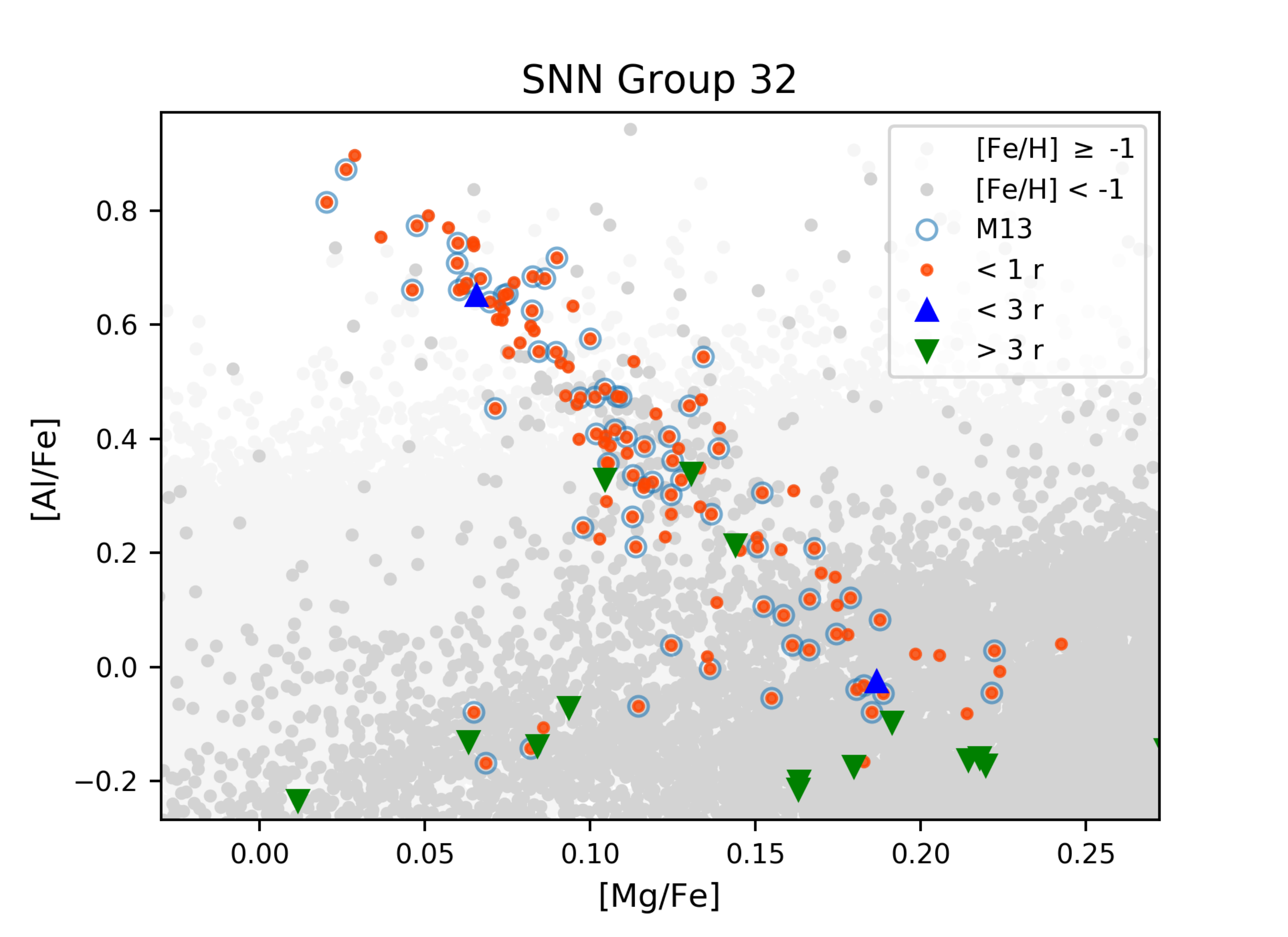}
  \caption{The [Al/Fe]-[Mg/Fe] anti-correlation is displayed for stars identified as part of group 32 by the SNN algorithm. The cluster identified 
in the chemistry-velocity space is M13. Stars identified by SNN within the
tidal radius are displayed (orange circles), along with confirmed members (circled orange points). Stars identified outside the tidal 
radius are potential candidates of a debris associate to the cluster (blue and green triangles). }
\label{fig:SNNf3}
\end{center}
\end{figure}

Fig.~\ref{fig:SNNf3} shows that stars of group 32 display an anti-correlation between [Al/Fe] and [Mg/Fe], which is a clear characteristic of globular clusters, even though our algorithm did not use [Al/Fe] and [Mg/Al]. A subsequent inspection shows that this clump of stars is indeed the globular cluster known as M13, found without a priori knowledge of its location in the sky. Similar to Fig.~\ref{fig:SNNf2}, the stars recovered by the algorithm and confirmed to be members are circled with open symbols in Fig.~\ref{fig:SNNf3}. Note that our algorithm identifies more stars than the known confirmed members. 

A study of these non--confirmed members shows evidence of anti-correlation between [Al/Fe] and [Mg/Fe]. It should be noticed, however, that while the 
majority of the non--confirmed members are within one tidal radius of the globular cluster (orange points), the remaining stars do follow the anti-correlation but are located outside the tidal radius. Two stars are between one and two tidal radii (blue triangles), and 15 stars are beyond three tidal radii away (green symbols). It is possible that these stars are candidate stellar debris of the globular cluster. Interestingly, even though our algorithm did not include [Al/Fe] and [Mg/Fe], we recovered the over-dense region in [Al/Fe] and [Mg/Fe] mentioned in Fig.~\ref{fig:AlMg}. Conversely, when we do include [Al/Fe] and [Mg/Fe] in our chemical tagging algorithms, our recovery rates never exceed 30\% for M13, much lower than the 95.8\% recovery rate from chemo-dynamical tagging.  

\begin{figure}[tb]
  \begin{center}
   \includegraphics[width=0.5 \textwidth]{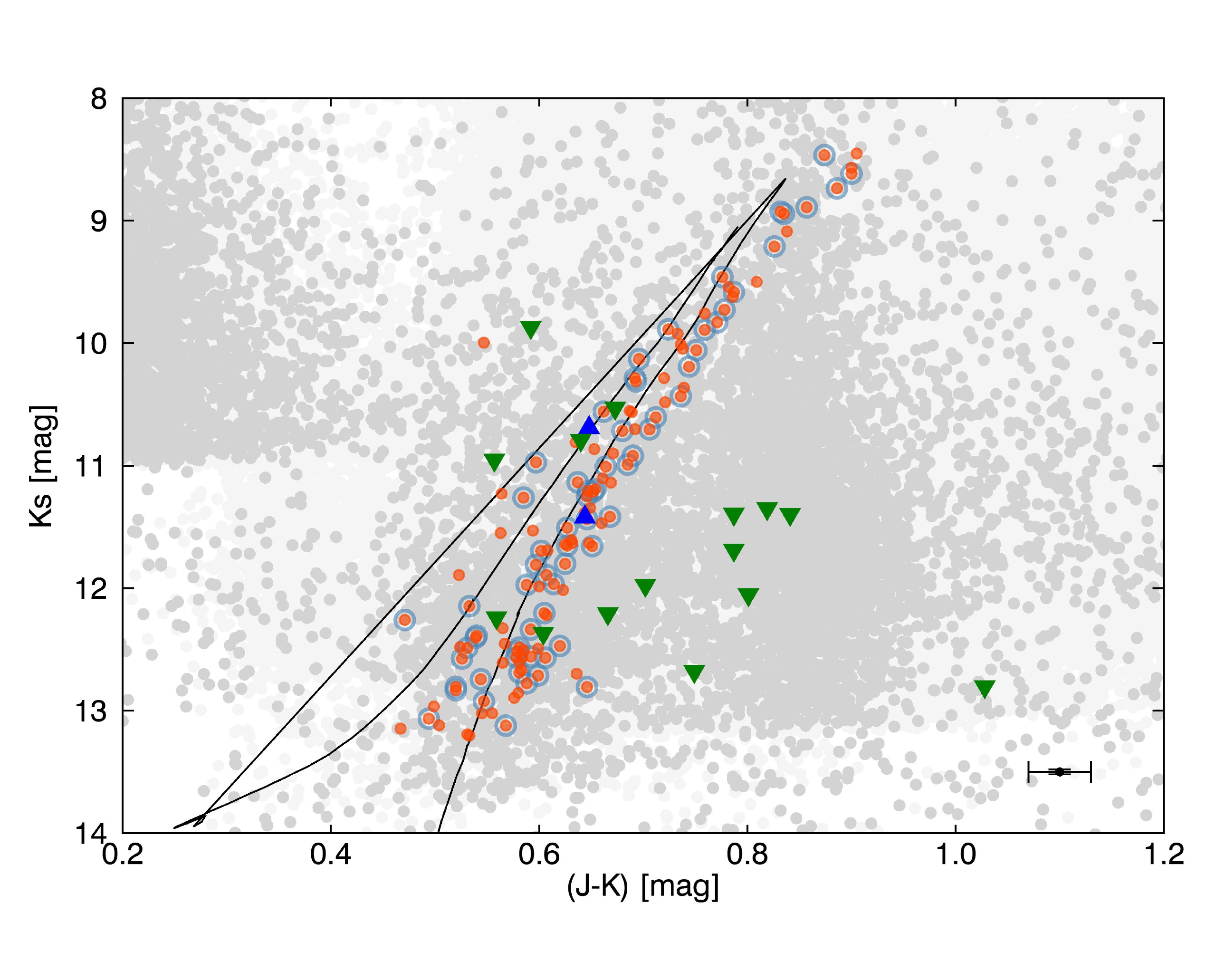}
  \caption{The colour-magnitude diagram of our M13 stars 
  together with a PARSEC isochrone derived from the parameters of \citet{OMalley17}. The meaning of symbols and colors is the same as in Fig.~\ref{fig:SNNf3}.}
\label{fig:CMD}
\end{center}
\end{figure}

As a next step we crossmatched the sample of stars obtained with our chemo-dynamical tagging algorithm with the 2MASS catalogue \citep{Skrutskie06}. Fig.~\ref{fig:CMD} displays the resulting photometry on a color-magnitude diagram (CMD) together with the PARSEC isochrone \citep{Bressan2012} of M13 calculated with the parameters of \citet{OMalley17} (E(B-V)=0.02, [Fe/H]=-1.58, (m-M)$_V$=14.54 mag and age=12.3 Gyr). The symbols and colors are the same as in Fig.~\ref{fig:SNNf3}. On the bottom-right of the plot we show the mean photometric errors of the sample. The 
distribution of the stars within one tidal radius on the CMD is in good 
agreement with the isochrone representing the cluster and the stars with distances up 
to 3 tidal radii also fall within the same distribution. In addition, 5 out of 15 stars at distances larger than 3 tidal radii are consistent with the infrared photometry of M13. While most outliers lie on the red side of the CMD, two stars are bluer than the mean (J-Ks) colour of the cluster, and one of them is situated within 
one tidal radius, 2M16414108+3627550. The unexpected color of this star 
is explained in the 2MASS archive: the J-band flux for this star is flagged 
as "detected at the location where the images were not deblended consistently 
in all three bands (JHKs)" and does not have any error listed in the 2MASS catalogue. Unfortunately, we could not perform a photometric analysis of the data at visible wavelength because the HST observations of M13, although available, constrain themselves at the very center of the cluster and thus exclude part of our sample.

\subsection{M3, NGC 5466, M5, and M92}
\label{sec:M92_M5}

We recovered 34.2\% of M3 members with a false positive rate of 34.9\% and 75\% of NGC 5466 members with a false positive rate of 14.3\%, each in their respective groups.

We did not recover M5, even though it has 103 members in our dataset. There are too many background stars with similar radial velocity to members of M5. There are 13,878 stars within the range of radial velocities of M5 members. Our choice of $N_{rv}=500$ is likely too low for M5 members to be the nearest neighbors of each other in the velocity space and in turn chemistry--velocity space.

We recovered known members from the M92 cluster but they were split into two separate groups. The two groups contain 26 and 10 members, respectively, with false positive rates of 44.7\% and 44.4\%. In Table \ref{tab:table1}, we show the recovery and false positive rates for the two groups combined. The split of M92 could be caused by the large number of background stars with similar radial velocities. There are 2,063 stars within the range of radial velocities of the M92 members. Again, our choice of $N_{rv}$ might be too low for the M92 members to be connect in velocity space.

When $N_{rv}$ is increased, we notice that different globular clusters start to merge with each other in our results. Since radial velocity is only one-dimensional, the data points are not able to spread out, as in the N-dimensional chemical space. We expect this to change when radial velocity is replaced with a higher dimensional velocity space.
	
\section{SUMMARY}
\label{sec:summary}

In this paper we have addressed the question whether stars with similar chemical
signatures can be identified as members of globular clusters. Our results
indicate that given the precision of the data in the measurements of the
chemical elements in the DR12 Cannon, the chemical properties have to be
combined with the stellar kinematic information, e.g. radial velocity, to
reproduce globular clusters in the sky. Our findings suggest that independent of the algorithm used,
clustering the data only with radial velocity or chemical abundances does not recover the membership of the globular clusters targeted in the APOGEE.  

We present a new algorithm, named Shared Nearest Neighbor Clustering (SNN), 
which is able to identify clusters in the chemistry-velocity space and we apply it to the DR12 Cannon data. We emphasize that the algorithm identifies clusters by taking the intersection of the nearest
neighbors in both chemical and radial velocity space. Unlike other clustering algorithms used in data mining, the SNN algorithm has the ability to find clusters of different sizes, shapes and densities and significantly increases the ability
to pick up members of globular clusters from among background
stars, given the accuracy of the data. In addition, it reduces the difficulty of choosing a single metric for a multi-dimensional space. 

As an example for the potential of our methodology, we showed that the stars
identified by the algorithm in one of our groups do indeed correspond to confirmed
members of the globular cluster M13. In addition, the stars identified
according to SNN anti-correlate in [Al/Fe]-[Mg/Fe], which is expected in
globular clusters. While the majority of the stars retrieved by the algorithm are located within one tidal-radius of M13, others following the same correlation are located outside and may be possible candidates of stellar debris. The color-magnitude diagram of the stars identified by our algorithm as cluster members supports the findings. We also learned that our algorithm is susceptible to the high number of background stars with similar properties.

\acknowledgments 
We thank the anonymous referee for constructive comments on this work. We are grateful to Juna Kollmeier for insightful comments. B.C. thanks Anru Zhang for the important discussion on statistical metrics. E.D. gratefully acknowledges the support of the ATP NASA Grant No NNX144AP53G. E.D., A.P., C.B.M., and E.K.G. acknowledge the support by 
Sonderforschungsbereich SFB 881 ``The Milky Way System'' (subprojects A8 and B5) of the German Research Foundation (DFG). \\

This project was developed in part at the 2017 Heidelberg Gaia Sprint, hosted by the Max-Planck-Institut f\"{u}r Astronomie, Heidelberg.\\

This publication makes use of data products from the Two Micron All Sky Survey, which is a joint project of the University of Massachusetts and the Infrared Processing and Analysis Center/California Institute of Technology, funded by the National Aeronautics and Space Administration and the National Science Foundation.\\

This project makes use of SDSS-III data. Funding for SDSS-III has been provided by the Alfred P. Sloan Foundation, the Participating Institutions, the National Science Foundation, and the U.S. Department of Energy Office of Science. The SDSS-III web site is http://www.sdss3.org/.

SDSS-III is managed by the Astrophysical Research Consortium for the Participating Institutions of the SDSS-III Collaboration including the University of Arizona, the Brazilian Participation Group, Brookhaven National Laboratory, Carnegie Mellon University, University of Florida, the French Participation Group, the German Participation Group, Harvard University, the Instituto de Astrofisica de Canarias, the Michigan State/Notre Dame/JINA Participation Group, Johns Hopkins University, Lawrence Berkeley National Laboratory, Max Planck Institute for Astrophysics, Max Planck Institute for Extraterrestrial Physics, New Mexico State University, New York University, Ohio State University, Pennsylvania State University, University of Portsmouth, Princeton University, the Spanish Participation Group, University of Tokyo, University of Utah, Vanderbilt University, University of Virginia, University of Washington, and Yale University.\\

\bibliographystyle{apj}
\bibliography{refs}

\end{document}